\documentclass[aps,prl,showpacs,superscriptaddress,groupedaddress]{revtex4}
\usepackage[latin9]{inputenc}
\usepackage{amsmath}
\usepackage{amssymb}
\usepackage{graphicx}
\usepackage{esint}

\makeatletter
\@ifundefined{textcolor}{}
{%
 \definecolor{BLACK}{gray}{0}
 \definecolor{WHITE}{gray}{1}
 \definecolor{RED}{rgb}{1,0,0}
 \definecolor{GREEN}{rgb}{0,1,0}
 \definecolor{BLUE}{rgb}{0,0,1}
 \definecolor{CYAN}{cmyk}{1,0,0,0}
 \definecolor{MAGENTA}{cmyk}{0,1,0,0}
 \definecolor{YELLOW}{cmyk}{0,0,1,0}
}

\newcommand{\lyxmathsym}[1]{\ifmmode\begingroup\def\b@ld{bold}
  \text{\ifx\math@version\b@ld\bfseries\fi#1}\endgroup\else#1\fi}

\@ifundefined{textcolor}{}{%
 \definecolor{BLACK}{gray}{0}
 \definecolor{WHITE}{gray}{1}
 \definecolor{RED}{rgb}{1,0,0}
 \definecolor{GREEN}{rgb}{0,1,0}
 \definecolor{BLUE}{rgb}{0,0,1}
 \definecolor{CYAN}{cmyk}{1,0,0,0}
 \definecolor{MAGENTA}{cmyk}{0,1,0,0}
 \definecolor{YELLOW}{cmyk}{0,0,1,0}
}

\usepackage{times}\usepackage{psfrag}\usepackage{epsf}\usepackage{epsfig}\usepackage{latexsym}\usepackage{amsfonts}\usepackage{color}\usepackage{epsf}\usepackage{latexsym}
 \usepackage{ifpdf}

\usepackage{dcolumn}\usepackage{bm}

\makeatother

\begin{document}

\title{Statics and dynamics of quasi one-dimensional Bose-Einstein condensate\\
in harmonic and dimple trap}

\author{Javed Akram}

\email{javedakram@daad-alumni.de}

\affiliation{ Institute für Theoretische Physik, Freie Universität Berlin, Arnimallee
14, 14195 Berlin, Germany}

\affiliation{Department of Physics, COMSATS, Institute of Information Technology
Islamabad, Pakistan}

\author{Axel Pelster}

\email{axel.pelster@physik.uni-kl.de}

\affiliation{ Fachbereich Physik und Forschungszentrum OPTIMAS, Technische Universität
Kaiserslautern, Germany}

\date{\today}
\begin{abstract}
We investigate a quasi one-dimensional $^{87}\text{Rb}$ Bose-Einstein
condensate in a harmonic trap with an additional dimple trap (dT)
in the center. Within a zero-temperature Gross-Pitaevskii mean-field
description we provide a one-dimensional physical intuitive model,
which we solve by both a time-independent variational approach and
numerical calculations. With this we obtain at first equilibrium results
for the emerging condensate wave function which reveal that a dimple
trap potential induces a bump or a dip in case of a red- or a blue-detuned
Gaussian laser beam, respectively. Afterwards, we investigate how
this dT induced bump/dip-imprint upon the condensate wave function
evolves for two quench scenarios. At first we consider the generic
case that the harmonic confinement is released. During the resulting
time-of-flight expansion it turns out that the dT induced bump in
the condensate wave function remains present, whereas the dip starts
decaying after a characteristic time scale which decreases with increasing
blue-detuned dT depth. Secondly, once the red- or blue-detuned dT
is switched off, we find that bright shock-waves or gray/dark bi-soliton
trains emerge which oscillate within the harmonic confinement with
a characteristic frequency. 
\end{abstract}

\pacs{05.30.Jp, 32.80.Pj, 03.75.Lm, 05.45.Yv,}

\maketitle

\section{Introduction}

The ability to manipulate and trap atoms with laser light has had
a tremendous development in many fields of physics. The very first
experimental success of trapping 500 Sodium atoms for several seconds
in the tight focus of a Gaussian red-detuned laser beam occurred in
1986 \cite{Chu}. The physical mechanism behind such an optical dipole
trap is the electric dipole interaction of the trapped polarized atoms
with the intense laser light, which is far detuned from the nearest
optical transition of the atoms. They are hence largely independent
from magnetic sublevels of the confined atoms, in contrast to a magneto-optical
trap (MOT) which can only trap atoms with a certain internal state
\cite{Grimm,Garrett}. The so called dimple trap (dT) is nothing but
a small tight optical dipole trap \cite{Ma,Comparat,Jacob}. Cooling
and trapping of atoms with these dT's has a strong impact on the study
of the Bose-Einstein condensates \cite{Stenger,Stellmer}, the observation
of long decay times for atoms in their ground state \cite{Davidson},
and the research of trapping other atomic species or molecules \cite{Weinstein}.

A straightforward method for realizing a dT is to rely on the potential
created by a freely propagating laser beam. The detuning of the laser
frequency versus the atomic resonances determines, whether the atoms
are red/blue-detuned, i.e. the laser frequency is below/above the
atomic resonance frequency, respectively \cite{Grimm}. The red-detuned
dT was particularly used for realizing matter wave traps \cite{Ashkin,Kurn,Ferez}
in the focus of a Gaussian laser beam. On the other hand, the blue-detuned
Gaussian laser beam was used in optical waveguides \cite{Xu,Song,Xu2,Noh,Bongs,Yin},
where the creation of repulsive potentials was demonstrated by using
Laguerre-Gaussian laser beam \cite{Gallatin,Soding,Kuga,Kuppens,Yin1,Webster}.
A focused or well-collimated Gaussian laser beam with a large red-detuning
\cite{Metcalf} or a dark hollow laser beam with a large blue-detuning
\cite{Yin2} were used to form 3D optical dipole trap's, which
can be widely applied to the accurate, non-contact manipulation and
control of cold atoms \cite{Barrett,Gustavson,Jacob}.

In this paper, we will focus on studying neutral $^{87}\text{Rb}$
atoms within a quasi one-dimensional harmonic trap with an additional
dimple trap. Experimentally, a highly elongated quasi-1D regime can
be reached by tightly confining the atoms in the radial direction,
effectively freezing-out the transverse dynamics \cite{Gorlitz,Moritz,Tolra,Hellweg,Kinoshita,Chuu,Hofferberth,Eckart,Pethick}.
It is worth mentioning that, when the transverse length scales are
of the order of or less than the atomic interaction length, the one-dimensional
system can only be described within the Tonks-Girardeau or within
the super-Tonks-Girardeau regime \cite{Olshanii,Petrov0,Bergeman},
which is experimentally realizable near a confinement-induced resonance
\cite{Paredes,Kinoshita0,Haller}. On the other hand, when the transverse
confinement is larger than the atomic interaction strength, the underlying
three-dimensional Gross-Pitaevskii equation (GPE) can be reduced to
an effective quasi 1D model \cite{Kamchatnov}. In one spatial dimension
(1D) this equation is well-known, for instance, to feature bright and dark solitons
for attractive and repulsive s-wave scattering lengths, respectively
\cite{Radouani,Kevrekidis,Frantzeskakis,Cuevas}. Many experiments
investigate the collision of two Bose-Einstein condensates where the
celebrated matter-wave interference pattern appears \cite{Andrews}
or shock-waves are generated \cite{Dutton}. For lower collisional
energies, the repulsive interaction energy becomes significant, and
the interference pattern evolves into an array of gray solitons \cite{Reinhardt,Kivshar,Scott,Busch,Ruostekoski,Khaykovich,Strecker,Shomroni}.
Furthermore, dark solitons can be created by manipulating the condensate
density using external potentials \cite{Burger,Denschlag,Carr,Becker}.

This work is organized as follows: In Sec.~II, we start with a model
which describes the dynamical evolution of a quasi-1D Bose-Einstein
condensate (BEC) in a magneto-optical trap with an additional red/blue-detuned
dimple trap in the center. Afterwards in Sec.~III, we justify a Thomas-Fermi
approximation for the condensate wave function and compare it with
numerical results. With this we show that the dT induces a bump or
a dip upon the condensate wave function depending on whether dT laser
beam is red- or blue-detuned. Subsequently, in Sec.~IV, we discuss
the dynamics of the dT induced bump/dip-imprint upon the condensate
wave function for two quench scenarios. After having released the
trap, the resulting time-of-flight expansion shows that the dT induced
imprint remains conserved for a red-detuned dT but decreases for a
blue-detuned dT. Furthermore, when the initial red/blue-detuned dT
is switched off, we observe the emergence of bright shock-waves or
gray/dark bi-soliton trains. Finally, Sec.~V summarizes our findings
for the proposed quasi-1D harmonically confined BEC with an additional
dimple trap in the center in view of a possible experimental realization.

\section{Quasi 1D model}

We start with the fact that the underlying Gross-Pitaevskii equation
for a condensate wave function can be formulated as the Hamilton principle
of least action with the action functional 
\begin{eqnarray*}
\mathcal{A}_{\text{3D}}=\int dt\int\mathcal{L_{\text{3D}}}\; d^{3}r,
\end{eqnarray*}
where the Lagrangian density reads for three spatial dimensions 
\begin{align}
\mathcal{L}_{\text{3D}}= & \frac{i\hbar}{2}\left[\psi^{\star}\left(\mathbf{r},t\right)\frac{\partial\psi\left(\mathbf{r},t\right)}{\partial t}-\psi\left(\mathbf{r},t\right)\frac{\partial\psi^{\star}\left(\mathbf{r},t\right)}{\partial t}\right]+\frac{\hbar^{2}}{2m_{\text{B}}}\psi^{\star}\left(\mathbf{r},t\right)\bigtriangleup\psi\left(\mathbf{r},t\right)-V(\mathbf{r})\psi^{\star}\left(\mathbf{r},t\right)\psi\left(\mathbf{r},t\right)\nonumber \\
 & -U_{\textrm{dT}}^{\textrm{3D}}\left(\mathbf{r}\right)\psi^{\star}\left(\mathbf{r},t\right)\psi\left(\mathbf{r},t\right)-\frac{G_{\text{B}}^{\text{3D}}}{2}\parallel\psi\left(\mathbf{r},t\right)\parallel^{4}\,.\label{eq1}
\end{align}
Here $\psi\left(\mathbf{r},t\right)$ describes the BEC wave function
with the coordinates $\mathbf{r}=\left(x,\; y,\; z\right)$ and the
two-particle interaction strength reads $G_{\text{B}}^{\text{3D}}=N_{\text{B}}4\pi\hbar^{2}a_{\text{B}}/m_{\text{B}}$,
where $N_{\text{B}}$ denotes the number of bosonic atoms. In case
of $^{87}\text{Rb}$ atoms, the s-wave scattering length is $a_{\text{B}}=94.7~{\rm {a}_{0}}$
with the Bohr radius ${\rm {a}_{0}}$. We assume that the bosons are
confined by a harmonic potential with an additional dT potential in
the center. 

For instance, a MOT yields a harmonic confinement $V\left(\mathbf{r}\right)=m_{\text{B}}\omega_{\text{z}}^{2}z^{2}/2+m_{\text{B}}\omega_{\text{r}}^{2}\left(x^{2}+y^{2}\right)/2$,
which has rotational symmetry with respect to the $z$-axis. In the
following, we consider the experimentally realistic trap frequencies
$\omega_{\text{r}}=2\pi\times160\,\textrm{Hz}\gg\omega_{\text{z}}=2\pi\times6.8\,\textrm{Hz}$
\cite{Garrett}, so we have a cigar-shaped condensate, where the oscillator
lengths amount to the values $l_{\text{r}}=0.84\,\mu\textrm{m}\ll l_{\text{z}}=4.12\,\mu\textrm{m}$.

An additional three-dimensional narrow Gaussian laser beam polarizes
the neutral atoms and, thus, yields the dT potential $U_{\textrm{dT}}^{\textrm{3D}}=U_{\textrm{0}}\textrm{I}\left(\mathbf{r}\right)$.
Within the rotating-wave approximation its amplitude $U_{\textrm{0}}$
is given by \cite{Milonni,Saleh,Allen,Scully} 
\begin{align}
U_{\textrm{0}} & =\frac{3\pi c^{2}\Gamma}{2\omega_{\textrm{A}}^{3}\Delta}\,.\label{eq:2}
\end{align}
Here $\Gamma=\left|<e|{\bf d}|g>\right|^{2}\omega_{\textrm{A}}^{3}/\left(3\pi\epsilon_{0}\hbar c^{3}\right)$
denotes the damping rate due to energy loss via radiation, which is
determined by the dipole matrix element between ground $g$ and excited
state $e$. Furthermore, $\Delta=\omega-\omega_{\textrm{A}}$ represents
the detuning of the laser, where $\omega$ is the laser frequency
and $\omega_{\textrm{A}}$ stands for the frequency of the D$_{1}$-
or D$_{2}$-line of $^{87}\text{Rb}$ atoms, which are the transitions
$5^{2}S{}_{1/2}\rightarrow5^{2}P{}_{1/2}$ or $5^{2}S{}_{1/2}\rightarrow5^{2}P{}_{3/2}$
with the wave lengths $\lambda_{\textrm{D}1}=794.76\,\textrm{nm}$ and
$\lambda_{\textrm{D}2}=780.03\,\textrm{nm}$, respectively. The intensity
profile of the Gaussian laser beam, which is assumed to move in $y$-direction,
is determined via 
\begin{align}
\textrm{I}\left(\mathbf{r}\right)=\frac{2P}{\pi W_{\text{x}}\left(y\right)W_{\text{z}}\left(y\right)}e^{-\left[\frac{2x^{2}}{W_{\text{x}}^{2}\left(y\right)}+\frac{2z^{2}}{W_{\text{z}}^{2}\left(y\right)}\right]}\,,
\end{align}
where $P$ denotes its power. Furthermore, $W_{\text{x/z}}(y)=W_{\text{0x/z}}\sqrt{1+y^{2}/y_{\text{Rx/z}}^{2}}$
defines the beam radius in $x$- and $z$-direction, where the intensity
decreases to $1/e^{2}$ of its peak value, and the Rayleigh lengths
$y_{\text{Rx/z}}=\pi W_{\text{0x/z}}/\lambda$ with wave length $\lambda=2\pi c/\omega$
define the distances where the beam radius increases by a factor of
$\sqrt{2}$ \cite{Saleh}. We use for the Gaussian laser beam width
along the $x$-axis $W_{\text{0x}}=1.1\,\mu\textrm{m}$ and along
the $z$-axis $W_{\text{0z}}=3.2\,\mu\textrm{m}$, which are about
ten times smaller than the corresponding ones used in Ref. \cite{Garrett}.
The corresponding Rayleigh lengths for the red-detuned laser light
with $\lambda=840\,\textrm{nm}$ \cite{Garrett} yield $y_{\text{Rx}}=4.526\,\mu\textrm{m}$
and $y_{\text{Rz}}=38.29\,\mu\textrm{m}$ and for the blue-detuned
laser light with $\lambda=772\,\textrm{nm}$ \cite{Xu-1} we get $y_{\text{Rx}}=4.92\,\mu\textrm{m}$
and $y_{\text{Rz}}=41.6\,\mu\textrm{m}$. Due to the fact $y_{\text{Rx}/\text{z}}\gg l_{\text{r}}$,
we can approximate the widths of the beam in $x$- and $z$-direction
according to $W_{\text{x}/\text{z}}(y)\approx W_{0\text{x}/\text{z}}$.
This simplifies the dimple trap to 
\begin{align}
U_{\textrm{dT}}^{\textrm{3D}}\left(\mathbf{r}\right) & =\frac{2U_{0}P}{\pi W_{0\text{x}}W_{0\text{z}}}e^{-\left(\frac{2x^{2}}{W_{\text{0x}}^{2}}+\frac{2z^{2}}{W_{\text{0z}}^{2}}\right)}.\label{eq:5-2}
\end{align}

As the MOT provides a quasi one-dimensional setting due to $a_{\text{B}}\ll l_{\text{r}}\ll l_{\text{z}}$,
we can follow Ref.~\cite{Kamchatnov}, and decompose the BEC wave-function
$\psi(\mathbf{r},t)=\psi(z,t)\phi({\bf r}_{\perp},t)$ with ${\bf r}_{\perp}=\left(x,\; y\right)$
and %
\begin{eqnarray}
\phi({\bf r}_{\perp},t) & = & \frac{e^{-\frac{x^{2}+y^{2}}{2l_{\text{r}}^{2}}}}{\sqrt{\pi}l_{\text{r}}}e^{-i\omega_{\text{r}}t}\,.\label{eq2}
\end{eqnarray}
Subsequently, we integrate out the two transversal dimensions of the
three-dimensional Lagrangian according to 
\begin{align}
\mathcal{L}_{\text{1D}}=\int_{-\infty}^{\infty}\int_{-\infty}^{\infty}\mathcal{L}_{\text{3D}}\; dxdy.
\end{align}
After a straight-forward calculation the resulting quasi one-dimensional
Lagrangian reads 
\begin{align}
\mathcal{L}_{\text{1D}}= & \frac{i\hbar}{2}\left[\psi^{\star}\left(z,t\right)\frac{\partial\psi\left(z,t\right)}{\partial t}-\psi\left(z,t\right)\frac{\partial\psi^{\star}\left(z,t\right)}{\partial t}\right]+\frac{\hbar^{2}}{2m_{\text{B}}}\psi^{\star}\left(z,t\right)\frac{\partial^{2}\psi\left(z,t\right)}{\partial z^{2}}-V(z)\psi^{\star}\left(z,t\right)\psi\left(z,t\right)\nonumber \\
 & -\textrm{U}e^{-\frac{2z^{2}}{W_{\text{0z}}^{2}}}\psi^{\star}\left(z,t\right)\psi\left(z,t\right)-\frac{G_{\text{B}}}{2}\parallel\psi\left(z,t\right)\parallel^{4},\label{eq4}
\end{align}
where $V\left(z\right)=m_{\text{B}}\omega_{z}^{2}z^{2}/2$ represents
an effective one-dimensional harmonic potential from the MOT, and
the one-dimensional two-particle interaction strength is 
\begin{equation}
G_{\text{B}}=2N_{\text{B}}a_{\text{B}}\hbar\omega_{\text{r}}\,.\label{eq:5}
\end{equation}
Furthermore, the one-dimensional dT depth turns out to be 
\begin{equation}
\textrm{U}=\frac{2U_{0}P}{\pi W_{\text{0z}}\sqrt{W_{\text{0x}}^{2}+2l_{\text{r}}^{2}}}\,,\label{eq:5-1}
\end{equation}
which depends on the power of the laser beam $P$ as well as via (\ref{eq:2})
and the detuning $\Delta=\omega-\omega_{\textrm{A}}$ on the laser
wave length $\lambda$. Note that the latter not only changes the
absolute value of the dT depth but also its sign via the detuning
$\Delta$. For red detuning, i.e. when the laser frequency is smaller
than the atomic frequency, the dT is negative and atoms are sucked
into the dT potential. In the opposite case of blue detuning the atoms
are repelled from the dT potential. Thus, the dT induces an imprint
on the BEC wave function, which can be either a bump for red detuning
or a dip for blue detuning. In the following we will analyze this
interesting effect in more detail.

To this end we consider the 1D action 
\begin{align}
\mathcal{A}_{\text{1D}}= & \int_{-\infty}^{\infty}\int_{-\infty}^{\infty}\mathcal{L}_{\text{1D}}\left(\psi^{\star}\left(z,t\right),\frac{\partial\psi^{\star}\left(z,t\right)}{\partial t},\frac{\partial\psi^{\star}\left(z,t\right)}{\partial z};\psi\left(z,t\right),\frac{\partial\psi\left(z,t\right)}{\partial t},\frac{\partial\psi\left(z,t\right)}{\partial z}\right)dzdt\label{eq8}
\end{align}
and determine the time dependent one-dimensional Gross-Pitaevskii
equation (1DGPE) according to the Euler-Lagrangian equation 
\begin{align}
\frac{\delta\mathcal{A}_{\text{1D}}\left[\psi^{\star},\psi\right]}{\delta\psi^{\star}\left(z,t\right)}= & \frac{\partial\mathcal{L}_{\text{1D}}}{\partial\psi^{\star}\left(z,t\right)}-\frac{\partial}{\partial z}\frac{\partial\mathcal{L}_{\text{1D}}}{\partial\frac{\partial\psi^{\star}\left(z,t\right)}{\partial z}}-\frac{\partial}{\partial t}\frac{\partial\mathcal{L}_{\text{1D}}}{\partial\frac{\partial\psi^{\star}\left(z,t\right)}{\partial t}}=0.\label{eq9}
\end{align}
By using the one-dimensional Lagrangian density (\ref{eq4}) the 1DGPE
reads 
\begin{equation}
i\hbar\frac{\partial}{\partial t}\psi(z,t)=\left\{ -\frac{\hbar^{2}}{2m_{B}}\frac{\partial^{2}}{\partial z^{2}}+\frac{m_{\text{B}}\omega_{z}^{2}}{2}z^{2}+\textrm{U}e^{-\frac{2z^{2}}{W_{\text{0z}}^{2}}}+G_{\text{B}}\parallel\psi(z,t)\parallel^{2}\right\} \psi(z,t).\label{eq10}
\end{equation}
On the right-hand side the first term represents the kinetic energy
of the atoms with mass $m_{\text{B}}$, the second term describes
the harmonic MOT potential, the third term stands for the dT potential,
and the last term represents the two-particle interaction. In order
to make Eq. (\ref{eq10}) dimensionless, we introduce the dimensionless
time as $\tilde{t}=\omega_{\text{z}}t$, the dimensionless coordinate
$\tilde{z}=z/l_{\text{z}}$, and the dimensionless wave function $\tilde{\psi}=\psi/\sqrt{l_{\text{z}}}$.
With this Eq. (\ref{eq10}) can be written in the form 
\begin{align}
i\frac{\partial}{\partial\tilde{t}}\tilde{\psi}\left(\tilde{z},\tilde{t}\right) & =\left\{ -\frac{1}{2}\frac{\partial^{2}}{\partial\tilde{z}^{2}}+\frac{1}{2}\tilde{z}^{2}+\tilde{\textrm{U}}e^{-\frac{\tilde{z}^{2}}{\tilde{\alpha}^{2}}}+\tilde{G}_{\text{B}}\parallel\tilde{\psi}\left(\tilde{z},\tilde{t}\right)\parallel^{2}\right\} \tilde{\psi}\left(\tilde{z},\tilde{t}\right),\label{eq11}
\end{align}
where we have $\tilde{G}_{\text{B}}=2N_{\text{B}}\omega_{\text{r}}a_{\text{B}}/\left(\omega_{\text{z}}l_{\text{z}}\right)$
and $\tilde{\textrm{U}}=\textrm{U}/\left(\hbar\omega_{\text{z}}\right)$.
For the above mentioned experimental parameters and $N_{\text{B}}=20\times10^{4}$
atoms of $^{87}\text{Rb}$, we obtain the dimensionless couplings
constant $\tilde{G}_{\text{B}}=11435.9$. Furthermore, the typical
dT depth $\left|\textrm{U}\right|/k_{\text{B}}=210\,\textrm{nK}$
yields the dimensionless value $\left|\tilde{U}\right|=643.83$, and
$\tilde{\alpha}=W_{\text{0z}}/\left(\sqrt{2}l_{\text{z}}\right)=0.548$
represents the ratio of the width of the dT potential along the $z$-axis
and the longitudinal harmonic oscillator length. From here on, we
will drop all tildes for simplicity.

\begin{figure}
\includegraphics[scale=0.9]{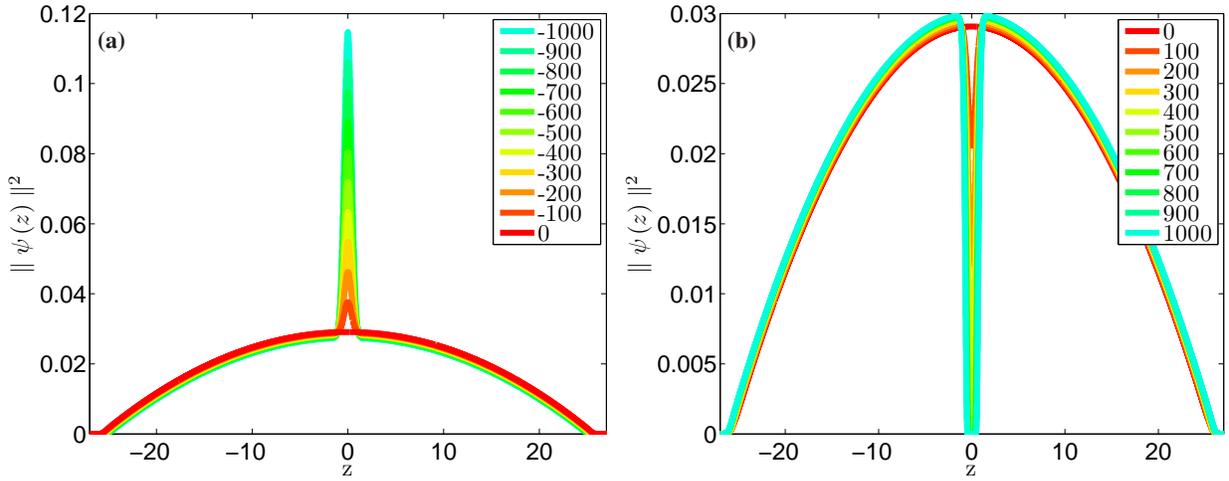}\\
 \caption{(Color online) Numerical density profile of BEC for the experimental
coupling constant value $G_{\text{B}}=11435.9$ and for the dT depth
U which increases from top to bottom according to the inlets. For
a) negative values of U, the bump in the condensate wave function
decreases, whereas for b) positive values the corresponding dip increases.
\label{Fig1}}
\end{figure}

\begin{figure}
\includegraphics[scale=0.4]{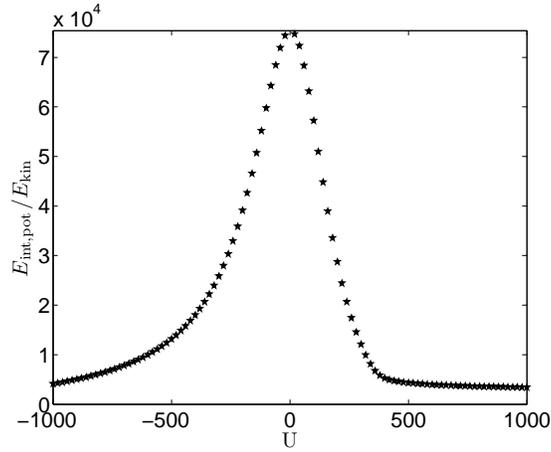} \caption{(Color online) Ratio $E_{\text{int,pot}}/E_{\text{kin}}$ versus U
from solving 1DGPE (\ref{eq11}). \label{Fig2}}
\end{figure}

\section{dT induced bump/dip-Imprint Upon Stationary Condensate Wave Function}

In order to determine the dT induced imprint on the condensate wave
function in equilibrium, we solve the 1DGPE (\ref{eq11}) in imaginary
time numerically by using the split-operator method \cite{Javanainen,Vudragovic}.
In this way we find that the dT-imprint leads to a bump/hole in the
BEC density at the trap center for negative/positive values of U as
shown in Fig.~\ref{Fig1}. For stronger red-detuned dT depth values
the bump increases further, but for stronger blue-detuned dT the dip
in the BEC density gets deeper and deeper until no more BEC atoms
remain in the trap center. After this qualitative overview on the
numerical results, we now work out an analytic approach for describing
this red/blue-detuned dT induced bump/dip on the BEC density in a more
quantitative way. To this end we present two arguments why the seminal
Thomas-Fermi (TF) approximation is also applicable in our context.

At first we provide a rough estimate in the case of an absent dT,
i.e. U$=0$, so the BEC density is characterized by the TF profile
$\psi\left(z\right)=\sqrt{\left(\mu-z^{2}/2\right)/G_{\text{B}}}\Theta\left(\mu-z^{2}/2\right)$,
where the Heaviside function $\Theta$ prevents the density to become
negative. Thus, the Tomas-Fermi radius $\sqrt{2\mu}$ follows from
the dimensionless chemical potential $\mu$, which is determined by
normalization to be $\mu=\frac{1}{2}\left(\frac{3}{2}\right)^{2/3}(G_{\text{B}})^{2/3}$.
As the red/blue-detuned dT is supposed to be inserted at the trap
center, we then calculate the dimensionless BEC coherence length $\xi$
at the trap center. It is defined by comparing the kinetic energy
$1/2\xi^{2}$ with the interaction energy in the trap center, which
is given by $\mu$. For the above mentioned experimental parameters
this yields the dimensionless BEC coherence length $\xi=0.038$, which
is about 14.4 times smaller than the dT width $\alpha=0.548$.
This indicates that the dT induced imprint upon the BEC wave-function
occurs on a length scale which is much larger than its coherence length,
so the TF approximation seems to be reasonable even in the presence
of the red/blue-detuned dT.

In view of a more quantitative justification for the applicability
of the Thomas-Fermi approximation, Fig.~\ref{Fig2} presents the
numerical result how the ratio of the sum of interaction and potential
energy $E_{\textrm{{\rm int,pot}}}$ versus the kinetic energy $E_{{\rm kin}}$
of the condensate wave function changes with increasing or decreasing
the red/blue-detuned dT depth U. The maximal value of this energy
ratio occurs for U$=0$ and amounts to 7.5$\times10^{4}$, which is
of the order of the number of particles. Furthermore, we read off
that the inequality $E_{\textrm{{\rm int,pot}}}/E_{\textrm{{\rm kin}}}\gg1$
holds within the whole region of interest for U, so the TF approximation
is there, indeed, valid.

\begin{figure}
\includegraphics[scale=0.5]{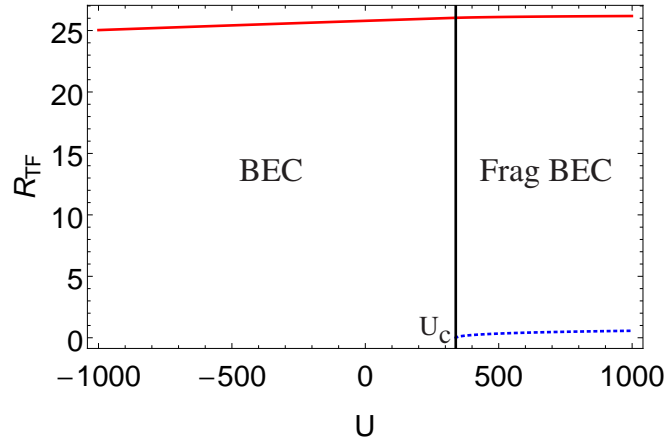} \caption{(Color online) Outer Thomas-Fermi radius $R_{\text{TF1}}$ (red solid)
and inner Thomas-Fermi radius $R_{\text{TF2}}$ (blue dashed) versus
dimple trap depth U. BEC fragments into two parts above $\textrm{U}_{\text{c}}\approx339.5$.
\label{Fig3} }
\end{figure}

Therefore, we investigate in the following the TF approximation in
more detail for non-zero red/blue-detuned dT depth U. To this end
we use for the condensate wave function the ansatz $\psi(z,t)=\psi(z)e^{-i\mu t}$,
insert it into the 1DGPE (\ref{eq11}), and neglect the kinetic energy
term, which yields the density profile 
\begin{eqnarray}
\psi\left(z\right)=\sqrt{\frac{1}{G_{\text{B}}}\left(\mu-\frac{z^{2}}{2}-\textrm{U}e^{-\frac{z^{2}}{\alpha^{2}}}\right)}\Theta\left(\mu-\frac{z^{2}}{2}-\textrm{U}e^{-\frac{z^{2}}{\alpha^{2}}}\right)\,.\label{eq13}
\end{eqnarray}
In view of the normalization $\intop_{-\infty}^{+\infty}\parallel\psi\left(z\right)\parallel^{2}dz=1$,
which fixes the chemical potential $\mu$, we have to determine the
Thomas-Fermi radii $R_{\text{TF}}$ from the condition that the condensate
wave function vanishes: 
\begin{eqnarray}
\mu=\frac{R_{\text{TF}}^{2}}{2}+\textrm{U}e^{-\frac{R_{\text{TF}}^{2}}{\alpha^{2}}}.\label{eq14}
\end{eqnarray}
As can be read off from Fig.~\ref{Fig1} the number of solutions
of Eq.~(\ref{eq14}) changes for increasing dT depth at a critical
value $\textrm{U}_{\text{c}}$, which follows from solving the implicit
equation 
\begin{eqnarray}
\textrm{U}_{\text{c}}=\frac{1}{2}\left(\frac{3}{2}\right)^{\frac{2}{3}}(G_{\text{B}}+\sqrt{\pi}\alpha\textrm{U}_{\text{c}})^{\frac{2}{3}}\,.\label{eq15}
\end{eqnarray}
This yields the result $\textrm{U}_{\text{c}}\approx339.5$ for the
experimental coupling constant $G_{\text{B}}=11435.9$, which compares
well with the value $\textrm{U}_{\text{c}}\approx342$ determined
from solving 1DGPE (\ref{eq11}). In the case that U is smaller than
$\textrm{U}_{\text{c}}$ Eq.~(\ref{eq14}) defines only the cloud
radius $R_{\text{TF1}}$. But for the case $\textrm{U}>\textrm{U}_{\text{c}}$
the dT drills a hole in the center of the $^{87}\text{Rb}$ condensate,
so it fragments into two parts. Thus, we have then apart from the
outer cloud radius $R_{\text{TF1}}$ also an inner cloud radius $R_{\text{TF2}}$.
With this the normalization condition $2\intop_{R_{\text{TF2}}}^{R_{\text{TF1}}}\parallel\psi\left(z\right)\parallel^{2}dz=1$
yields 
\begin{eqnarray}
\mu\left(R_{\text{TF1}}-R_{\text{TF2}}\right)-\frac{1}{6}\left(R_{\text{TF1}}^{3}-R_{\text{TF2}}^{3}\right)=\frac{G_{\text{B}}}{2}+\frac{\sqrt{\pi}\alpha\textrm{U}}{2}\left[\text{Erf}\left(\frac{R_{\text{TF1}}}{\alpha}\right)-\text{Erf}\left(\frac{R_{\text{TF2}}}{\alpha}\right)\right]\,,\label{eq16}
\end{eqnarray}
where $\text{Erf}(y)=\frac{2}{\sqrt{\pi}}\int_{0}^{y}e^{-x^{2}}dx$
denotes the error function. In case of $\textrm{U}\le\textrm{U}_{\text{c}}$
the inner cloud radius $R_{\text{TF2}}$ vanishes and the cloud radius
is approximated via $R_{\text{TF1}}\approx\sqrt{2\mu}$ due to Eq.
(\ref{eq14}) as it is much larger than the dimple trap width $\alpha$. Thus, the chemical potential is determined explicitly from
\begin{eqnarray}
\mu\approx\frac{1}{2}\left(\frac{3}{2}\right)^{2/3}\left(G_{\text{B}}+\sqrt{\pi}\alpha\textrm{U}\right)^{2/3}\,,\quad\textrm{U}\le\textrm{U}_{\text{c}}\,.\label{eq17}
\end{eqnarray}
Provided that $\textrm{U}\ge\textrm{U}_{\text{c}}$, the inner cloud
radius $R_{\text{TF2}}$ has to be taken into account according to
Fig.~\ref{Fig1} and, due to the fact that $R_{\text{TF2}}^{2}\ll\textrm{U}$,
we get from Eq.~(\ref{eq14}) the approximation $\mu\approx\textrm{U}e^{-\frac{R_{\text{TF2}}^{2}}{\alpha^{2}}}$,
which reduces to 
\begin{eqnarray}
R_{\text{TF2}}\approx\alpha\sqrt{\log\left(\frac{\textrm{U}}{\mu}\right)}.\label{eq18}
\end{eqnarray}
Thus, we conclude that $R_{\text{TF2}}$ vanishes, indeed, at $\textrm{U}_{\text{c}}$
according to Eq.~(\ref{eq15}) and Eq.~(\ref{eq17}). With this
we obtain from Eq.~(\ref{eq16}) that the chemical potential follows
from solving 
\begin{eqnarray}
3\left[\sqrt{\pi}\alpha\textrm{U}+G_{\text{B}}+2\alpha\mu\sqrt{\log\left(\frac{\textrm{U}}{\mu}\right)}\right]\approx3\sqrt{\pi}\alpha\textrm{U}\text{Erf}\left(\sqrt{\log\left(\frac{\textrm{U}}{\mu}\right)}\right)+\alpha^{3}\log^{\frac{3}{2}}\left(\frac{\textrm{U}}{\mu}\right)+4\sqrt{2}\mu^{3/2}\,,\,\textrm{U}\ge\textrm{U}_{\text{c}}\,.
\end{eqnarray}
Figure \ref{Fig3} shows the resulting outer and inner Thomas-Fermi
radius as a function of the dT depth U. We read off that $R_{\text{TF1}}\approx\sqrt{2\mu}$
remains approximately constant for $\textrm{U}\ge\textrm{U}_{\text{c}}$,
so we conclude that the chemical potential $\mu$ is locked to its
critical value $\mu_{c}\approx\textrm{U}_{\text{c}}=339.5$. Furthermore,
we note that the inner Thomas-Fermi radius $R_{\text{TF2}}$ increases
up to about $5.4\alpha$ for the considered range of $\textrm{U}$.

\begin{figure}
\includegraphics[scale=0.4]{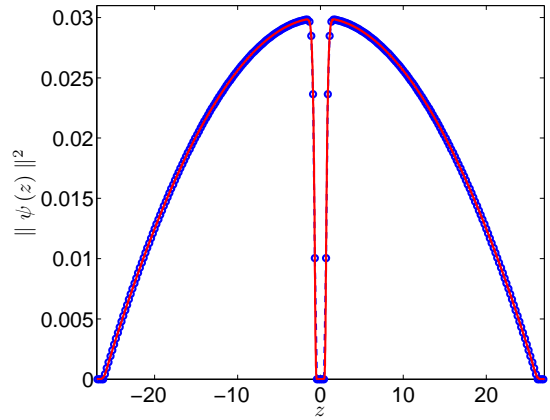} \caption{(Color online) Condensate density for $\textrm{U}=1000$ from solving
1DGPE (\ref{eq11}) in imaginary time numerically (red) and from TF
approximation (\ref{eq13}) (blue-circles). \label{Fig4}}
\end{figure}

\begin{figure}[th]
\includegraphics[scale=0.9]{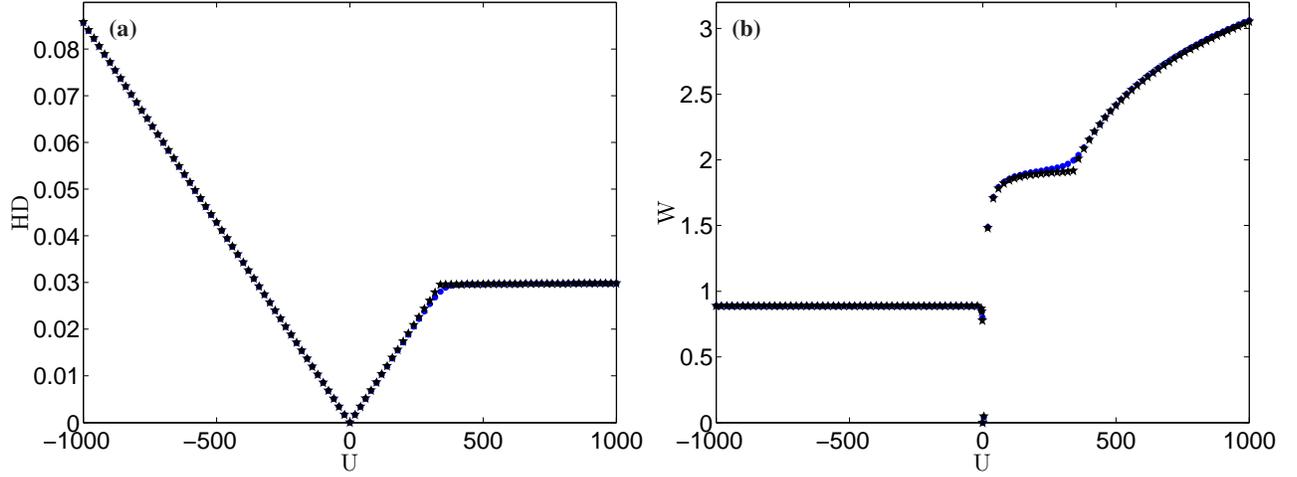} \caption{(Color online) a) Height/depth and b) width of the dT induced bump/dip
according to Eqs.~(\ref{eq20})--(\ref{eq21}), versus the red/blue-detuned
dT depth U for the experimental BEC coupling constant $G_{\text{B}}=11435.9$
calculated numerically by solving 1DGPE (\ref{eq11}) in imaginary
time (blue circles) and analytically (black stars) from the TF condensate
wave function (\ref{eq13}). \label{Fig5}}
\end{figure}

Figure \ref{Fig4} compares the resulting TF condensate wave function
(\ref{eq13}) with a numerical solution of the 1DGPE (\ref{eq11})
in imaginary time at U$=1000$ and we read off that both agree quite
well. Thus, our TF approximation describes the equilibrium properties
of the condensate wave function in the presence of the red/blue-detuned
dT even quantitatively correct. In view of a more detailed comparison,
we characterize the red/blue-detuned dT induced imprint upon the condensate
wave function $\psi(z)$ by the following two quantities. The first
one is the hight/depth (HD) of the dT induced imprint 
\begin{eqnarray}
 & \begin{array}{r@{}l@{\,}l}
\text{HD}=\left\{ \begin{array}{r@{}l@{\,}l}
\parallel\psi\left(0\right)\parallel_{\textrm{U}}^{2}-\parallel\psi\left(0\right)\parallel_{\textrm{U}=0}^{2}\qquad\textrm{U}\leq0\\
\,\\
\text{Max}\left(\parallel\psi\left(z\right)\parallel_{\textrm{U}}^{2}\right)-\parallel\psi\left(0\right)\parallel_{\textrm{U}}^{2}\qquad\textrm{U}\geq0
\end{array}\right.\end{array}\label{eq20}
\end{eqnarray}
and the second one is the red/blue-detuned dT induced imprint width
$\text{W}$, which we define as follows. For $\textrm{U}\leq0$ we
use the full width half maximum 
\begin{eqnarray}
\parallel\psi\left(\text{W/2}\right)\parallel_{\textrm{U}}^{2}=\left(\parallel\psi\left(0\right)\parallel_{\textrm{U}}^{2}+\parallel\psi\left(0\right)\parallel_{\textrm{U}=0}^{2}\right)/2\qquad\textrm{U}\leq0\,,
\end{eqnarray}
whereas for $\textrm{U}>0$ we define the equivalent width \cite{Carroll}:
\begin{eqnarray}
\text{W}=\left(2I_{\text{0}}z_{\text{Max}}-\intop_{-z_{\text{Max}}}^{z_{\text{Max}}}\parallel\psi\left(z\right)\parallel_{\textrm{U}}^{2}dz\right)/\left(I_{\text{0}}-\parallel\psi\left(0\right)\parallel_{\textrm{U}}^{2}\right)\qquad\textrm{U}>0\,,\label{eq21}
\end{eqnarray}
where we have $I_{\text{0}}=\text{Max}\left(\parallel\psi\left(z\right)\parallel_{\textrm{U}}^{2}\right)$.
Figure~\ref{Fig5} a) shows the red/blue-detuned dT induced imprint
height/depth as a function of U. At first, we read off that for $\textrm{U}=0$,
i.e.~when we have not switched on the dT, bump/dip vanishes. Furthermore,
in the range $\textrm{U}\leq\textrm{U}_{\text{c}}$ we observe that
height/depth of the dT induced imprint bump/dip changes linearly with
U according to 
\begin{eqnarray}
\text{HD}\approx\frac{|\textrm{U}|}{G_{\text{B}}}\,.\label{eq22}
\end{eqnarray}
In case of $\textrm{U}>\textrm{U}_{\text{c}}$ height/depth of the
dT induced imprint has approximately the constant value $\text{HDc}=\textrm{U}_{\text{c}}/G_{\text{B}}\approx0.029$
as follows from the TF wave function (\ref{eq13}) and the above mentioned
locking of the chemical potential to its critical value. Note that
this constant value only slightly deviates from the corresponding
numerical value $\text{HDc}=0.03$.

Correspondingly, Figure~\ref{Fig5} b) depicts the dimple trap induced
width W as a function of U. From our TF approximation we obtain for
the width transcendental formulas, which read in case of $\textrm{U}\leq0$
\begin{eqnarray}
\frac{\text{W}^{2}}{4}+2\textrm{U}e^{-\frac{\text{W}^{2}}{4\alpha^{2}}}-\textrm{U}+\frac{1}{2}\left(\frac{3}{2}\right)^{2/3}\left[\left(G_{\text{B}}\right)^{2/3}-\left(G_{\text{B}}+\sqrt{\pi}\alpha\textrm{U}\right)^{2/3}\right]=0\,,\label{eq23}
\end{eqnarray}
and for $\textrm{U}>\alpha^2/2$ 
\begin{eqnarray}
\text{W}=\frac{2\alpha^{3}\sqrt{\log\left(\frac{2\textrm{U}}{\alpha^{2}}\right)}\left[2\log\left(\frac{2\textrm{U}}{\alpha^{2}}\right)+3\right]-6\sqrt{\pi}\alpha\textrm{U}\text{Erf}\left(\sqrt{\log\left(\frac{2\textrm{U}}{\alpha^{2}}\right)}\right)}{3\left[\alpha^{2}+\alpha^{2}\log\left(\frac{2\textrm{U}}{\alpha^{2}}\right)-2\textrm{U}\right]}\,.\label{eq24}
\end{eqnarray}
As shown in Fig.~\ref{Fig5} b), for an increasing red-detuned dT
depth, the width remains approximately constant, but just before $\textrm{U}=0$
starts to decrease to zero. For a blue-detuned dT the width of the
dip continuously increases with an intermediate plateau at $\textrm{U}_{\text{c}}$
with the value $\text{Wc}\approx1.91$, which agrees well with the
numerically obtained one $\text{Wc}\approx1.99$.

\section{Dimple trap induced imprint upon Condensate Dynamics}

In an experiment, any dT induced bump/dip upon the condensate wave
function could only be detected dynamically. Thus, it is of high interest
to study theoretically whether the dT induced bump/dip imprint, which
we have found and analyzed for the stationary case in the previous
section, remains present also during the dynamical evolution of the
condensate wave function. To this end we investigate two quench scenarios
numerically in more detail. The first one is the standard time-of-flight
(TOF) expansion after having switched off the harmonic trap when the
amplitude of the red/blue-detuned dT is still present. In the second
case we consider the inverted situation that the red/blue-detuned
dT is suddenly switched off within a remaining harmonic confinement,
which turns out to give rise to the emergence of bright shock-waves
or bi-solitons trains, respectively.

\subsection{Time-of-Flight Expansion}

Time-of-flight (TOF) absorption pictures represent an important diagnostic
tool to analyze dilute quantum gases since the field's inception.
By suddenly turning off the magnetic trap, the atom cloud expands
with a dynamics which is determined by both the momentum distribution
of the atoms at the instance, when the confining potential is switched
off, and by inter-atomic interactions \cite{Mewes,Inouye}. We have
investigated the time-of-flight expansion dynamics of the BEC with
the dT by solving numerically the 1DGPE (\ref{eq11}) and analyzing
the resulting evolution of the condensate wave function. It turns
out that, despite the continuous broadening of the condensate density,
its dT induced imprint remains qualitatively preserved both for red
and blue-detuned dT. Therefore, we focus a more quantitative discussion
upon the dynamics of the corresponding dT induced imprint height/depth
and width.

\begin{figure}
\includegraphics[scale=0.9]{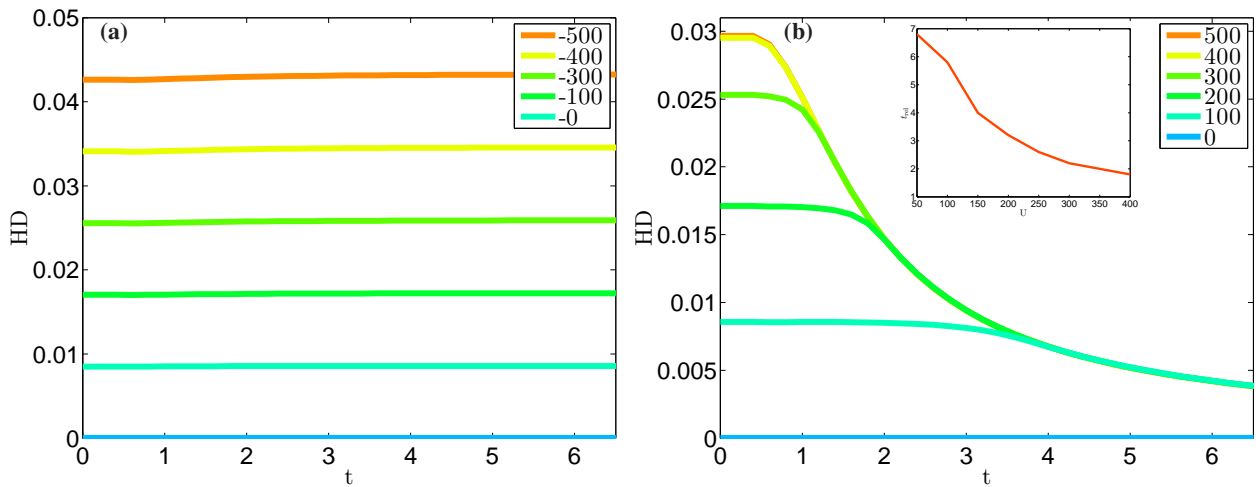} \caption{(Color online) Height/depth of the dT induced imprint after having
released the harmonic trap versus time for a) increasing negative
and b) decreasing positive values of dT depth U from top to bottom.
Inlet: relaxation time $t_{\text{rel}}$ decreases with increasing
U. \label{Fig6} }
\end{figure}

For a red-detuned dT, it turns out that the bump height even
remains constant in time. This is shown explicitly in Fig.~\ref{Fig6}
a), which roughly preserves its initial value at $t=0$. In case of
the bump width, we even find that no significant changes do
occur neither in time nor for varying U, therefore we do not present
a corresponding figure. Note that the latter finding originates from
Fig.~\ref{Fig5} b), where the width is shown to be roughly constant
for all dT depths.

\begin{figure}
\includegraphics[scale=0.5]{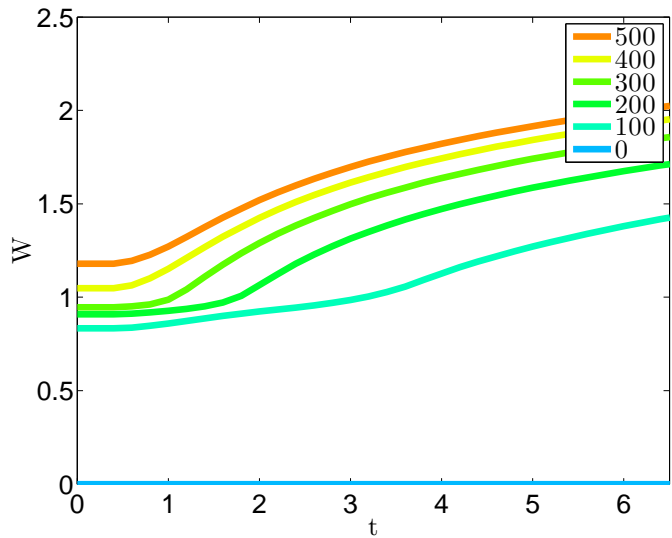} \caption{(Color online) Width of the dT induced imprint after having released
the trap versus time for decreasing positive values of dT depth U
from top to bottom. \label{Fig7} }
\end{figure}

Instead, in case of a blue-detuned dT, the dip decays after a characteristic
time scale as shown in Fig.~\ref{Fig6} b). The inlet reveals that
the dip relaxes with a shorter time scale for increasing blue-detuned
dT depth U. In addition, we read off from Fig.~\ref{Fig7} that,
at the beginning of TOF, the dT induced imprint width remains at first
constant and then increases gradually. This change of W occurs on
the scale of the relaxation time of HD, which is depicted in Fig.~\ref{Fig6}
b).

\subsection{Wave Packets Versus Solitons}

\begin{figure}
\includegraphics{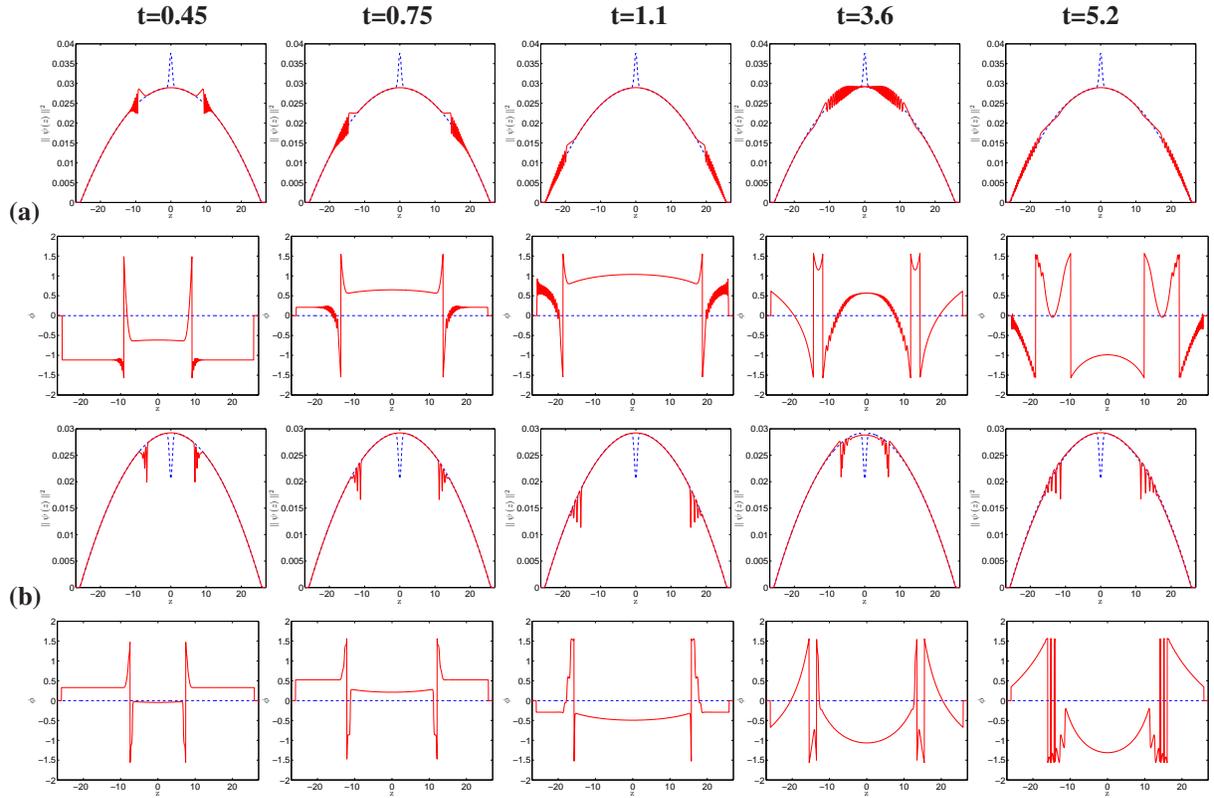} \caption{(Color online) Density (phase) profile of BEC after having switched
off the red/blue-detuned dT: in blue-dashed line at $t$=0, and in
red-solid line at $t$=0.45 ($1^{{\rm st}}$ column), $t$=0.75 ($2^{{\rm nd}}$
column), $t$=1.1 ($3^{{\rm rd}}$ column), $t$=3.6 ($4^{{\rm th}}$
column), and $t$=5.2 ($5^{{\rm th}}$ column) for (a) $U$=-100 and
(b) $U$=100. \label{Fig8}}
\end{figure}

\begin{figure}
\includegraphics[scale=0.8]{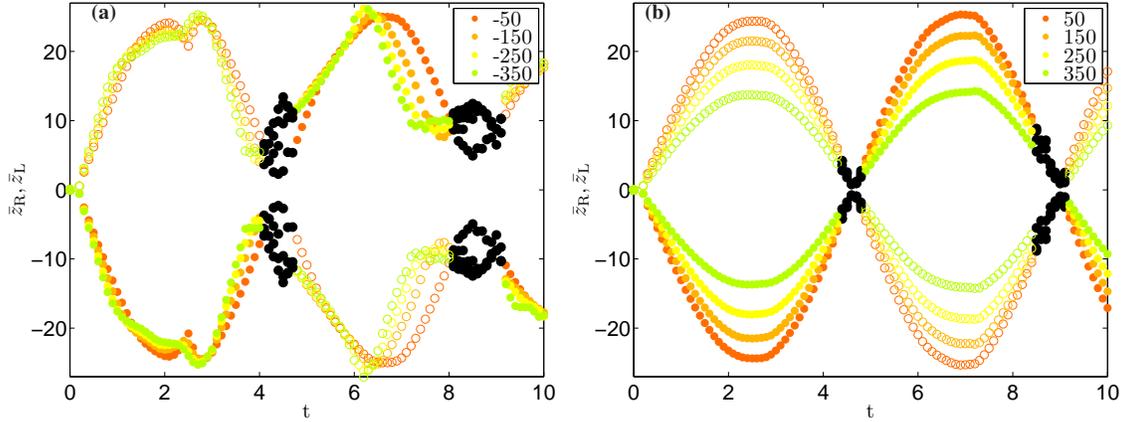} \caption{(Color online) Center of mass positions of excitations $\bar{z}_{{\rm L}}$
(filled circles) and $\bar{z}_{{\rm R}}$ (empty circles) according
to Eq. (\ref{eq25}) versus time after having switched off the dT
with increasing absolute value of the depth $|\textrm{U}|$ from top
to bottom, for a) red-detuning and b) blue-detuning. Black filled
circles represent the region of colliding excitations, where mean
positions are not perfectly detectable. \label{Fig9}}
\end{figure}

Due to their quantum coherence, BECs exhibit rich and complex dynamic
patterns, which range from the celebrated matter-wave interference
of two colliding condensates \cite{Andrews,Dutton} over Faraday waves
\cite{engels,balaz} to the particle-like excitations of solitons
\cite{Reinhardt,Kivshar,Scott,Busch,Ruostekoski,Becker,Shomroni}.
For our 1D model of a BEC with a harmonic and a dimple trap in the
center, we investigated the dynamics of the condensate wave function
which emerges after having switched off the dT. To this end, Fig.~\ref{Fig8}
depicts the resulting profile of density $n=|\psi|^{2}$ and phase
$\phi=\tan^{-1}\left(\psi_{{\rm Re}}/\psi_{{\rm Im}}\right)$ of the
condensate wave function $\psi$ at different instants of time. Both
for an initial red- and blue-detuned dT, we observe that two excitations
of the condensate are created at the dT position, which travel in
opposite direction with the same center-of-mass speed, are reflected
at the trap boundaries and then collide at the dT position. Furthermore,
we find that these excitations qualitatively preserve their shape
despite the collision and that the BEC wave function reveals characteristic
phase slips between $-\pi/2$ and $\pi/2$. All these findings are
not yet conclusive to decide whether these excitations represent wave
packets in the absence of dispersion or solitons. Therefore, we investigate
their dynamics in more detail, by determining their center-of-mass
motion via 
\begin{equation}
\bar{z}_{{\rm {L,R}}}\left(t\right)=\frac{\int_{-\infty,0}^{0,\infty}z\left(\parallel\psi\left(z,t\right)\parallel_{\text{U}}^{2}-\parallel\psi\left(z,t\right)\parallel_{\text{U}=0}^{2}\right)dz}{\int_{-\infty,0}^{0,\infty}\left(\parallel\psi\left(z,t\right)\parallel_{\text{U}}^{2}-\parallel\psi\left(z,t\right)\parallel_{\text{U}=0}^{2}\right)dz}\,,\label{eq25}
\end{equation}
which are plotted in Fig.~\ref{Fig9}. Note that the mean positions
$\bar{z}_{{\rm {L}}}$ and $\bar{z}_{{\rm {R}}}$ of the excitations
are uncertain in the region where they collide. Nevertheless Fig.~\ref{Fig9}
demonstrates that the excitations oscillate with the frequency $\Omega=2\pi\times4.87~{\rm Hz}$
irrespective of sign and size of U. As we have assumed the trap frequency
$\omega_{{\rm z}}=2\pi\times6.8~{\rm Hz}$, we obtain the ratio $\Omega/\omega_{{\rm z}}\approx0.72$,
which is quite close to $\Omega/\omega_{{\rm z}}=1/\sqrt{2}\approx0.707$. 

Despite these similarities of the cases of an initial red and blue-detuned
dT, we observe one significant difference. Whereas the oscillation
amplitudes of the excitations do not depend on the value of the initial
$\textrm{U}<0$ according to Fig.~\ref{Fig9} a), we find decreasing
oscillation amplitudes of the excitations with increasing the initial
$\textrm{U}>0$ in Fig.~\ref{Fig9} b). Such an amplitude dependence
on the initial condition is characteristic for gray/dark solitons
according to Ref.~\cite{Busch}. This particle-like interpretation
of the excitations agrees with the other theoretical prediction of
Ref.~\cite{Busch} that gray/dark solitons oscillate in a harmonic
confinement with the frequency $\Omega/\omega_{{\rm z}}=1/\sqrt{2}$,
which was already confirmed in the Hamburg experiment of Ref.~\cite{Becker}
and is also seen in Fig.~\ref{Fig9}. 

Conversely, for an initial red-detuned dT the excitations can not
be identified with bright solitons as the dynamics is governed by a GPE with
a repulsive two-particle interaction. Here the excitations have to
be interpreted as wave packets which move without any dispersion as
follows from a Bogoliubov dispersion relation and the smallness of
the coherence length. Thus, for $\textrm{U}<0$ the excitations
propagate like sound waves in the BEC \cite{andrews1} and, within
a TF approximation, their center-of-mass motion is described by the evolution
equation \cite{Damski1} 
\begin{eqnarray}
\frac{dz(t)}{dt}=\sqrt{\mu-\frac{z^{2}(t)}{2}}\,.\label{EV}
\end{eqnarray}
Solving (\ref{EV}) with the initial condition $z(0)=0$ yields the
result $z(t)=\sqrt{2\mu}\sin t/\sqrt{2}$. Thus, we read off that
the oscillation amplitude coincides with the TF radius and that the
dimensionless oscillation frequency turns out to be $\Omega=1/\sqrt{2}$
in agreement with Fig.~\ref{Fig9} a). 

Thus, we conclude that switching off the red/blue-detuned dT leads
to physically different situations. For an initial red-detuned dT,
we generate wave packets which correspond to white shock waves \cite{Damski2},
whereas for the corresponding blue-detuned case bi-soliton trains
emerge \cite{Reinhardt,Strecker,Khawaja}, due to the collision of
the two fragmented parts of the condensate. Note that it can be shown
in our system that gray bi-solitons trains are generated for a partially
fragmented BEC, i.e. $\textrm{U}<\textrm{U}_{\text{c}}$. On the other
hand the dark bi-solitons trains turn out to be only generated for
$\textrm{U}\geq\textrm{U}_{\text{c}}$, where the BEC is well fragmented
into two parts equilibrium.

\section{Summary and Conclusion }

In the present work we studied within a quasi 1D model both analytically
and numerically how a dimple trap in the center of a harmonically
trapped BEC affects the condensate wave function. At first, we showed
for the equilibrium properties of the system that the Thomas-Fermi
approximation agrees quantitatively with numerical solutions of the
underlying 1D Gross-Pitaevskii equation. For an increasing red-detuned
dT depth, it turns out for the induced bump that the height decreases
linearly, whereas the width remains approximately constant. In contrast
to that we found for an increasing blue-detuned dT that depth and
width of the induced dip initially increase. Beyond a critical value
$\textrm{U}_{\text{c}}$, the BEC even fragments into two parts and,
if U is increased beyond $\textrm{U}_{\text{c}}$, the dT induced
imprint yields a condensate wave function whose width increases further,
although the dip height/depth remains constant. Afterwards, we investigated
the dT induced bump/dip upon the condensate dynamics for two quench
scenarios.

At first, we considered the release of the harmonic confinement, which
leads to a time-of-flight expansion and found that the dT induced
imprint remains conserved for a red-detuned dT but decreases in the
blue-detuned case. This result suggests that it might be experimentally
easier to observe the bump for a red-detuned dT. On the other hand,
in an experiment one has to take into account that inelastic collisions
lead to two- and three-body losses of the condensate atoms \cite{Roberts,Carretero}.
As such inelastic collisions are enhanced for a higher BEC density,
they play a vital role for a red-detuned dT, when the condensate density
has a bump at the dT position, but are negligible for the blue-detuned
dT with the dip in the BEC wave-function. Thus, a more realistic description
of the experiment needs to consider the loss of condensate atoms by
adding damping terms to the 1DGPE (\ref{eq11}), which are of the
form $i\Upsilon_{2}\parallel\tilde{\Psi}\left(\tilde{z}\right)\parallel^{2}$
and $i\Upsilon_{3}\parallel\tilde{\Psi}\left(\tilde{z}\right)\parallel^{4}$,
where the positive constants $\Upsilon_{2}$ and $\Upsilon_{3}$ denote
two- and three-body loss rates, respectively. We note that these additional
terms may have nontrivial effects on the dT properties \cite{Bao}.

In addition, we analyzed the condensate dynamics after having switching
off the red/blue-detuned dT. This case turned out to be an interesting
laboratory in order to study the physical similarities and differences
of bright shock-waves and gray/dark bi-soliton trains, which emerge for an initial red- and blue-detuned dT, respectively.
The astonishing observation, that the oscillation frequencies of both
the bright shock-waves and the bi-soliton trains coincide, is presumably
an artifact of the harmonic confinement. Thus, it might be rewarding
to further investigate these different dynamical features also in
anharmonic confinements \cite{Dalibard,Kling1,Kling2}. Additionally,
we have also found that the generation of gray/dark bi-soliton trains
is a generic phenomenon on collisions of partially/fully fragmented
BEC, respectively, and the partially/fully fragmented BEC is strongly
depending upon the equilibrium values of the dimple trap depth.

\section{Acknowledgment}

We thank James Anglin, Antun Bala\v{z}, Thomas Bush, Herwig Ott, Ednilson
Santos, and Artur Widera for insightful comments. Furthermore, we
gratefully acknowledge financial support from the German Academic
Exchange Service (DAAD). This work was also supported in part by the
German-Brazilian DAAD-CAPES program under the project name ``Dynamics
of Bose-Einstein Condensates Induced by Modulation of System Parameters\textquotedblright{}
and by the German Research Foundation (DFG) via the Collaborative
Research Center SFB/TR49 ``Condensed Matter Systems with Variable
Many-Body Interactions\textquotedblright{}.

\end{document}